\documentclass[sigconf]{acmart}
\AtBeginDocument{%
  \providecommand\BibTeX{{%
    \normalfont B\kern-0.5em{\scshape i\kern-0.25em b}\kern-0.8em\TeX}}}

\copyrightyear{2021}
\acmYear{2021}
\setcopyright{acmcopyright}\acmConference[MM '21]{Proceedings of the 29th ACM International Conference on Multimedia}{October 20--24, 2021}{Virtual Event, China}
\acmBooktitle{Proceedings of the 29th ACM International Conference on Multimedia (MM '21), October 20--24, 2021, Virtual Event, China}
\acmPrice{15.00}
\acmDOI{10.1145/3474085.3475625}
\acmISBN{978-1-4503-8651-7/21/10}

\usepackage{multirow}
\usepackage{bm}

\usepackage[belowskip=2pt,aboveskip=0pt]{caption}

\begin{document}
\fancyhead{} 
\title{Disentangling Hate in Online Memes}


\author{Roy Ka-Wei Lee}
\email{roy\_lee@sutd.edu.sg}
\affiliation{%
  \institution{Singapore University of Design and Technology}
  \country{Singapore}
  \city{Singapore}
 }

\author{Rui Cao}
\authornote{Both authors contributed chiefly and equally to this research. The author order is incorrect but not revisable due to conference policy.}
\email{ruicao.2020@phdcs.smu.edu.sg}
\affiliation{%
  \institution{Singapore Management University}
  \country{Singapore}
  \city{Singapore}
 }

\author{Ziqing Fan}
\authornotemark[1]
\email{ziqingfan0331@gmail.com}
\affiliation{%
  \institution{University of Electronic Science and Technology of China}
  \city{Chengdu}
  \country{China}
 }

\author{Jing Jiang}
\email{jingjiang@smu.edu.sg}
\affiliation{%
  \institution{Singapore Management University}
  \country{Singapore}
  \city{Singapore}
 }
 
\author{Wen-Haw Chong}
\email{whchong.2013@phdis.smu.edu.sg}
\affiliation{%
  \institution{Singapore Management University}
  \country{Singapore}
  \city{Singapore}
 }

\renewcommand{\shortauthors}{Cao and Fan, et al.}

\begin{abstract}
Hateful and offensive content detection has been extensively explored in a single modality such as text. However, such toxic information could also be communicated via multimodal content such as online memes. Therefore, detecting multimodal hateful content has recently garnered much attention in academic and industry research communities. This paper aims to contribute to this emerging research topic by proposing \textsf{DisMultiHate}, which is a novel framework that performed the classification of multimodal hateful content. Specifically, \textsf{DisMultiHate} is designed to disentangle  target entities in multimodal memes to improve the hateful content classification and explainability. We conduct extensive experiments on two publicly available hateful and offensive memes datasets. Our experiment results show that \textsf{DisMultiHate} is able to outperform state-of-the-art unimodal and multimodal baselines in the hateful meme classification task. 

\end{abstract}




\maketitle

{\color{red} \textbf{Disclaimer}: \textit{This paper contains violence and discriminatory content that may be disturbing to some readers. Specifically, Figures 1 and 2 and Tables 6 and 7 contain actual examples of hateful memes and hate speech targeting particular groups. These examples are very offensive and distasteful. However, we have made the hard decision to display these actual hateful examples to provide context on the toxicity of malicious content that we are dealing with. Besides making technical contributions in this paper, we hope the distasteful examples used could also raise awareness of the vulnerable groups targeted in hate speeches in the real-world.}}

\section{Introduction}
The proliferation of social media has enabled users to share and spread ideas at a prodigious rate. While the information exchanges in social media platforms may improve an individual’s sense of connectedness with real and virtual communities, these platforms are increasingly exploited for the propagation of hateful content that attacks or uses discriminatory languages targeting a person or a group based on their race, religion, gender, etc. \cite{fortuna2018survey,schmidt2017survey}. The spread of online hate speech has sowed discord among individuals or communities online and resulted in violent hate crimes. Therefore, it is a pressing issue to detect and curb online hateful content.

The existing research on automated hateful content detection has predominantly focused on text-based content \cite{fortuna2018survey,schmidt2017survey}, neglecting multimodal content such as memes. \textit{Memes}, which are images with short texts, are a popular communication form in online social media. Although the memes are typically humorous in nature, they are also increasingly used to spread hate. These hateful memes often target certain communities and or individuals based on race, religion, gender, or physical attributes, by portraying them in a derogatory manner~\cite{kiela2020hateful,suryawanshi2020multimodal,gomez2020exploring}. The hateful memes could also be a greater threat to social peace than text-based online hate speeches due to their viral nature; users could re-post or share these hateful memes in multiple conversations and contexts. 

\begin{figure}[t] 
	\centering
	\setlength{\tabcolsep}{0pt} 
	\renewcommand{\arraystretch}{0} 
	\begin{tabular}{ccc}
		\includegraphics[scale = 0.3]{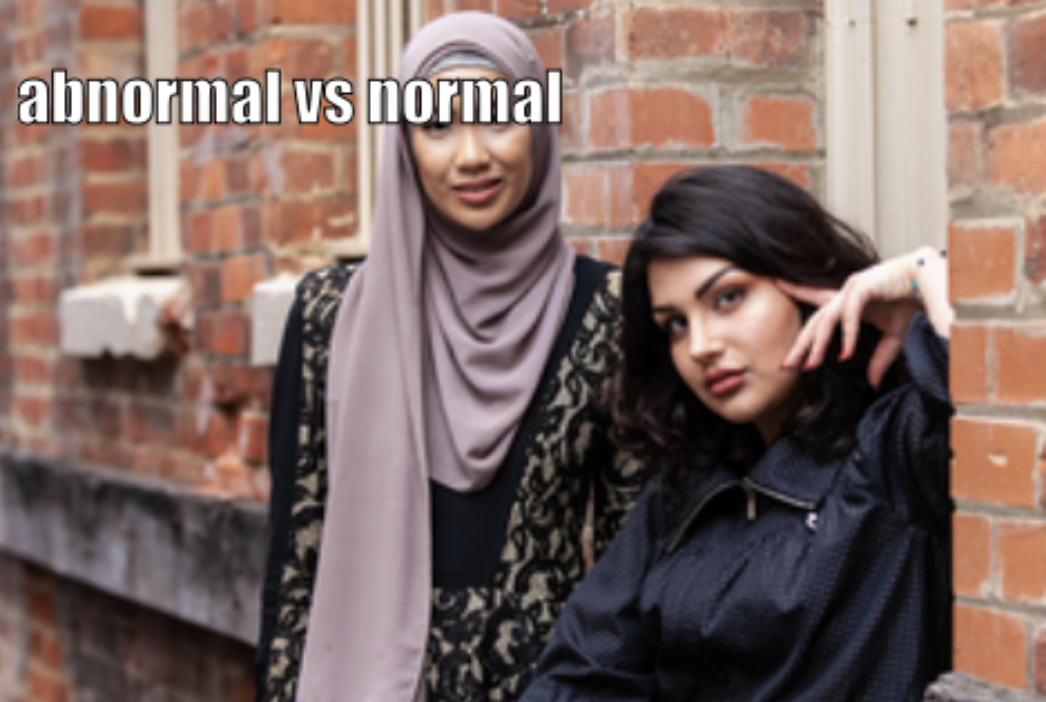} 
	\end{tabular}
	\caption{Example of hateful meme.}
	\label{fig:exmaple}
\end{figure}

To combat the spread of hateful memes, social networking platforms such as Facebook had recently released a large hateful meme dataset as part of a challenge to encourage researchers to submit solutions to perform hateful memes classification~\cite{kiela2020hateful}. The research community has answered this challenge by exploring and proposing multimodal classification models \cite{lippe2020multimodal,velioglu2020detecting,muennighoff2020vilio,zhang2020hateful,das2020detecting,zhou2020multimodal}. Several studies had proposed solutions that focused on innovating the fusion techniques to combine text and visual features extracted from memes to perform the classification tasks~\cite{suryawanshi2020multimodal,kiela2020hateful}. Others have explored fine-tuning large-scale pre-trained multimodal methods to perform hateful memes classification~\cite{kiela2020hateful,lippe2020multimodal,velioglu2020detecting,muennighoff2020vilio,zhang2020hateful,das2020detecting,zhou2020multimodal}. Nevertheless, these existing methods have limited explainability and cannot reason the context embedded in the hateful memes.

Hateful meme classification is a challenging task. Consider an example of a hateful meme illustrated in Figure~\ref{fig:exmaple}; when we examine the text and image as independent features, the content seems normal and benign. However, when we interpret the meme as a whole, the underlying message is very offensive and hateful. Another key element that helped us understand hateful meme is the context illustrated in the memes, and often the target entities (e.g., race, religion, etc.) in the hateful message provide important contextual information. For the example in Figure~\ref{fig:exmaple}, the target entity of the hateful content would be both gender and religion (i.e., female Muslim). Existing multimodal hateful meme classification models are unable to capture such target entity contextual information.

This paper aims to address the research gaps by proposing a novel framework, \textsf{DisMultiHate}, which learns and disentangles the representations of hate speech-related target entities, such as race and gender, in memes to improve the hateful content classification. 
Our framework includes a novel self-supervising training task that enables us to extract the target entities using disentangled latent representations. The disentangled representations serve as contextual information to improve hateful meme classification. 

We summarize this paper's contribution\footnote{Code: https://gitlab.com/bottle\_shop/safe/dismultihate} as follows: (i) We propose a novel multimodal hateful meme classification model called \textsf{DisMultiHate}, which disentangles the representations of hate speech-related target entities to improve performance and explainability of hateful meme classification. (ii) We conduct extensive experiments on two publicly available datasets. Our experiment results show that \textsf{DisMultiHate} consistently outperforms state-of-the-art methods in the hateful meme classification. (iii) We conduct case studies and demonstrated \textsf{DisMultiHate}'s ability to identify target entities in hateful memes, thereby providing contextual explanations for the classification results.

\section{Related Work}


\subsection{Hate Speech Detection}
With the proliferation of social media and social platforms, automatic detection of hate speech has received considerable attention from the data mining, information retrieval, and natural language processing (NLP) research communities. Several text-based hate speech detection datasets~\cite{waseem2016hateful,ParkF17,DavidsonWMW17} have been released. Previous works exploit both machine learning based methods~\cite{xiang2012detecting,chen2012detecting,waseem2016you,nobata2016abusive,chatzakou2017mean} and deep learning based techniques to tackle the problem~\cite{cao2020deephate,djuric2015hate,mehdad2016characters,gamback2017using,badjatiya2017deep,grondahl2018all,zhang2018detecting}. The existing automated hate speech detection method has yielded good performance. However, most of the existing studies have focused on text-based hateful content, neglecting the rich multimedia user-generated content.


\subsection{Multimodal Hate Speech}
To address the gap in hate speech detection research, recent works have attempted to  explore multimodal hateful content classification tasks such as detecting online hateful memes~\cite{kiela2020hateful,lippe2020multimodal,velioglu2020detecting,muennighoff2020vilio,zhang2020hateful,das2020detecting,zhou2020multimodal}. The flourish of multimodal hateful meme detection studies could be attributed to the availability of several hateful memes datasets published in recent year~\cite{kiela2020hateful,suryawanshi2020multimodal,gomez2020exploring}. For instance, Facebook had proposed the \textit{Hateful Memes Challenge}\footnote{https://ai.facebook.com/blog/hateful-memes-challenge-and-data-set/}, which encouraged researchers to submit solutions to perform hateful memes classification~\cite{kiela2020hateful}. A dataset consists of 10K memes were published as part of the challenge, and the memes are specially constructed such that unimodal methods cannot yield good performance in this classification task. Therefore, existing studies have motivated to adopted a multimodal approach to perform hateful memes classification.

The existing multimodal hateful memes classification approaches can be broadly categorized into two groups: (a) models that adopt early fusion techniques to concatenate text and visual features for classification~\cite{suryawanshi2020multimodal,kiela2020hateful}, and (b) models that directly fine-tune large scale pre-trained multimodal models~\cite{kiela2020hateful,lippe2020multimodal,velioglu2020detecting,muennighoff2020vilio,zhang2020hateful,das2020detecting}. Recent studies have also attempted to use data augmentation~\cite{zhou2020multimodal,zhu2020enhance} and ensemble methods~\cite{sandulescu2020detecting,velioglu2020detecting} to enhance the hateful memes classification performance. Nevertheless, these existing methods have limited explainability and cannot reason the context embedded in the hateful memes. For example, hate speech should involve abusive content targeted at an individual or a group~\cite{awal2021angrybert,DavidsonWMW17,cambridge}. However, most hateful memes classification methods cannot identify the hate targets and unable to provide the context for the hateful content. 

This paper aims to address the research gaps by proposing a novel framework, \textsf{DisMultiHate}, which learns and disentangles the representations of hate speech-related target entities, such as race and gender, in memes to improve the hateful content classification. Unlike \cite{zhu2020enhance}, which augment general entities information as input, we extract the target entities using disentangled latent representations learned using self-supervised training. The disentangled representations serve as contextual information to provide some form of explanation to the hateful meme classification.

\subsection{Disentangling Representation Learning}
\citet{bengio2013representation} defined a disentangled representation as one where single latent units are sensitive to changes in a single generative factor while being relatively invariant to changes in other factors. Hence, \cite{bengio2013representation} proposed to factorize and learn disentangled representations by training independent units to encode different aspects of input data. Various methods have also been proposed to perform representation disentanglement. For instance, existing studies have attempted to learn disentangled representations using supervised signals~\cite{hinton2011transforming,karaletsos2015bayesian,goroshin2015learning}. Other works have explored purely unsupervised approaches to disentangled factor learning by exploiting the information-theoretic aspect of the variational auto-encoders and adding a regularization term that minimizes the mutual information between different units of the representations~\cite{burgess2018understanding,chen2018isolating,HigginsMPBGBML17}.

Latent representation disentanglement has been applied in various tasks. For example, representation disentanglement had been applied to learn and disentangle users' preferences for different categories of items to improve recommendation~\cite{ma2020disentangled,ma2019learning}. \citet{bouchacourt2018multi} had exploited the disentangled representation learning to group data based on disentangled visual semantics. \citet{dupont2018learning} had disentangled representations to learn factors that correspond to handwriting numbers' various characteristics. This paper contributes the state-of-the-art in disentangled representation learning by disentangling the representation of target entities in online hateful memes. To the best of our knowledge, this is the first work that applies representation disentanglement on the hateful memes classification task.

\section{Preliminaries}
\subsection{Problem Definition}
We define the problem of hateful memes multimodal classification as follows: Given an image $\mathcal{I}$ and a piece of text $\mathcal{O}$ consisting of a sequence of words, a classification model will predict the label of the multimodal meme (\textit{hateful} or \textit{non-hateful}). This binary classification task can also be regarded as a regression task, where a model predicts a confidence score $y \in \mathbb{R}$ ranging from zero to one, indicating the likelihood of the meme being hateful. Specifically, the meme would be regarded as hateful if the predicted score is above a threshold $\lambda$; otherwise, the meme is predicted to be non-hateful.

\subsection{Datasets}
\label{sec:dataset}
We train and evaluate our proposed model on two popular and publicly-available hateful datasets: \textit{Facebook hateful memes (FHM} and \textit{MultiOff}. Table \ref{tab:dataset} shows the distributions of the datasets.


\begin{table}[t]
  \caption{Distributions of FHM and MultiOFF datasets}
  \label{tab:dataset}
  \begin{tabular}{c|p{6em}|p{6em}|p{6em}}
    \hline
    \textbf{Dataset}& \textbf{train} &\textbf{validation} & \textbf{test}\\\hline\hline\hline
    FHM & hateful (3050), non-hateful (5450) & hateful (250), non-hateful (250) & hateful (500), non-hateful (500)  \\\hline\hline
    MultiOFF & offensive (187), non-offensive (258) &  offensive (58), non-offensive (91) & offensive (58), non-offensive (91)\\\hline\hline
\end{tabular}
\end{table}

\textbf{Facebook Hateful Memes (FHM)}~\cite{kiela2020hateful}:
The dataset was constructed and released by Facebook as part of a challenge to crowd-source multimodal hateful meme classification solutions. The dataset contains $10K$ memes with binary labels(i.e., hateful or non-hateful). The Facebook challenge did not release the labels of the memes in the test split. Therefore, we utilize the \textit{dev-seen} split as the \textit{test}. Existing studies have also adopted the same dataset setting in their training and evaluations~\cite{lippe2020multimodal,zhou2020multimodal,zhu2020enhance}.

\textbf{MultiOFF}~\cite{suryawanshi2020multimodal}: The dataset contains $1K$ memes related to the 2016 United States presidential election. The memes are labeled as \textit{offensive} or \textit{non-offensive}. Although the authors have labeled the memes based on offensiveness, we manually examined the memes and found that the offensive memes could also be considered hateful as most of them contain abusive and discriminatory messages against a person or minority group.

\subsection{Data Pre-processing}
\label{sec:preprocessing}
To improve our proposed method's reproducibility, we also provide an overview of the data pre-processing steps applied on the FHM and MultiOFF memes datasets. Specifically, the following steps were taken to pre-process the memes in the datasets:

\begin{itemize}
    \item \textbf{Image resizing}: The datasets provided memes in all sizes. We resized the images proportionally to a minimum of 140 pixels and a maximum of 850 pixels. This ensures consistency of the visual input into our proposed model and baselines.
    \item \textbf{Text extraction and removal}: We extract and remove the text in the memes using open-source Python packages EasyOCR\footnote{https://github.com/JaidedAI/EasyOCR} and MMEditing\footnote{https://github.com/open-mmlab/mmediting}. The texts are removed from the memes to facilitate better entity and demographic detection.
    \item \textbf{Entity detection}: To augment the memes with relevant external information, we leverage Google Vision Web Entity Detection API\footnote{https://cloud.google.com/vision/docs/detecting-web} to detect and caption the entities in the cleaned image. The detected entities provide contextual information on the memes. 
    \item \textbf{Demographic detection}: Often, hate speeches are targeted at groups based on demographic information such as race and gender, and such information servers as important contextual information in hateful meme classification. To augment the memes with demographic information, we utilized the FairFace classifier~\cite{karkkainen2019fairface} to detect and classify the faces in the images, then mapping the label back to the person's bounding box with the largest overlapped area with the face. Noted that the demographic information is only extracted when the meme contains human entities.
\end{itemize}

The pre-processed datasets serve as input for the training of our proposed model discussed in the next section.

\section{Proposed Model}

\begin{figure*}[t] 
	\centering
	\setlength{\tabcolsep}{0pt} 
	\renewcommand{\arraystretch}{0} 
	\begin{tabular}{ccc}
		\includegraphics[scale = 0.5]{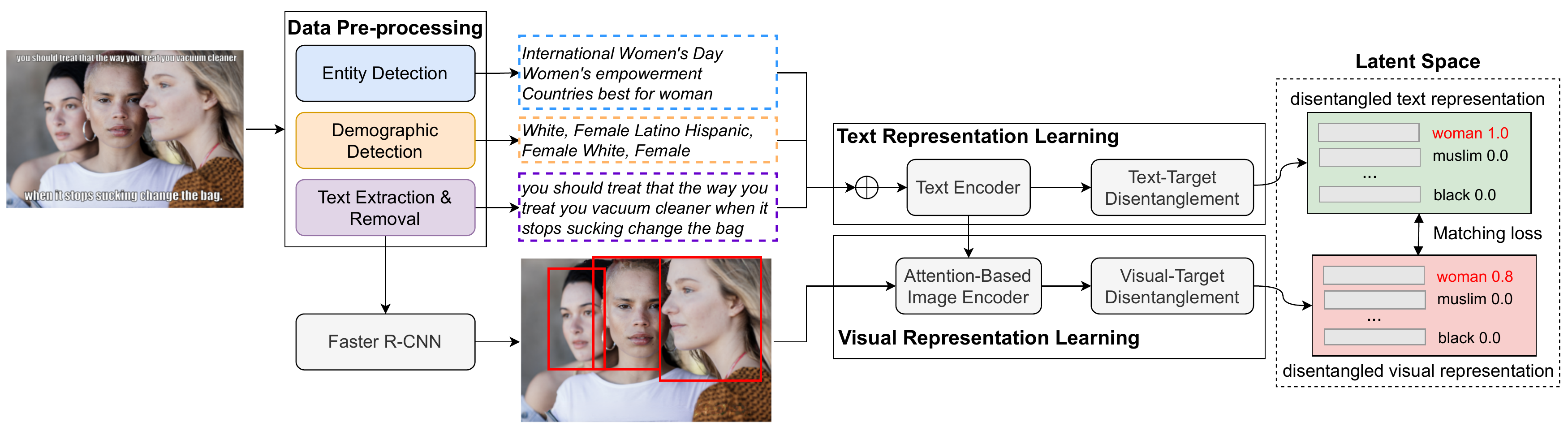} 
	\end{tabular}
	\caption{Architecture of our \textsf{DisMultiHate} model.}
	\label{fig:arch}
\end{figure*}


Figure~\ref{fig:arch} illustrates the architectural framework of our proposed \textsf{DisMultiHate} model. Broadly, \textsf{DisMultiHate} consists of three main modules: (a) \textit{data pre-processing}, (b) \textit{text representation learning}, and (c) \textit{visual representation learning}. The details of the data pre-processing module are discussed in Section~\ref{sec:preprocessing}. The goal of the text representation learning module is to learn a disentangled latent representation of the combined textual information output from the data pre-processing module. The details of the text representation learning module will be discussed in Section~\ref{sec:text}. 
The visual representation learning module aims to learn a disentangled latent representation based on the meme's image. The details of the visual representation learning module will be discussed in Section~\ref{sec:visual}. 

A core element in the two learning modules is the process of disentangling the target information from the text and visual representations. Specifically, the latent text and visual representations are projected into a disentangled latent space $\mathcal{D}$, where each latent unit of the disentangled representation represents a probability for a certain category of hate (i.e., religion, gender and race, etc.). For a multimodal meme, disentangled targets from the textual modality and the visual modality should be consistent with each other. Therefore, we introduce a self-supervised matching loss, which constrains the disentangled visual representation to be similar to the disentangled text representation. Finally, the learned text and visual representations will be fed into a regression layer to predict the likelihood of the meme being hateful. The classification process will be discussed in Section \ref{sec:classification}.


\subsection{Text Representation Learning}
\label{sec:text}
This module is designed to learn a disentangled latent representation of the textual information extracted from a meme. The input of the module is the concatenation of the text information output of the data pre-processing step. Specifically, we concatenated the text extracted from the meme, detected entities, and demographic information. Formally, we denote the concatenated text information as $\mathbf{O}=\{\mathbf{o}_j\}_{j=1}^M$, where $\mathbf{o}_j \in \mathbb{R}^{|\mathcal{V}|}$ is the one-hot vector representation for the $j$-th word in the text's word sequence, $M$ is the length of the text, and $\mathcal{V}$ is the vocabulary.



\textbf{Text Encoder}. The concatenated text information $\mathbf{O}$ is first fed into a text encoder to generate latent text representations. Since the input text involves words from various domains such as religion, politics, or military, a powerful text encoder is required to capture the semantics in textual information. Bidirectional Encoder Representations from Transformers (BERT)~\cite{devlin2018bert}, which has demonstrated its superiority in various natural language processing (NLP) tasks, is an ideal text encoder for our task. We initialize the BERT with pre-trained weights and fine-tune it with our task. Using the BERT text encoder, we generate the textual representations as follow:
\begin{equation}
    [\mathbf{s}, \mathbf{C}] = \text{BERT}([\mathbf{w}_\text{[CLS]},\mathbf{O}]),
\end{equation}
where $\mathbf{w}_\text{[CLS]}\in \mathbb{R}^{|\mathcal{V}|}$ denotes the one-hot representations for the ``[CLS]'' token, $[\cdot,\cdot]$ is the concatenation operation and $\mathbf{C}=\{\mathbf{c}_j\}_{j=1}^{M}$ is the set of textual representations, and $\mathbf{c}_j \in \mathbb{R}^u$ is the representation for the $j$-th word in the input text $\mathbf{O}$. Similar to ~\cite{devlin2018bert}, we utilize the representation of the ``[CLS]'' token as the sentence representation $\mathbf{s} \in \mathbb{R}^u$.

\textbf{Text-target disentanglement}. The latent text representation generated using the BERT encoder captures rich information on the meme's semantics. However, for our hateful meme classification task, we are interested in contextual information present in the latent text representation, specifically the targets of the hateful content. For example, in Figure~\ref{fig:arch}, we aim to identify ``gender'' as the target entity of the hateful message. Therefore, we need to design a mechanism to disentangle the target information in the latent text representation $\mathbf{s}$. 


We first transform the the sentence representation $\mathbf{s}$ into a latent space using a linear projection layer:
\begin{equation}
  \mathbf{s}_\text{p} =\mathbf{W}_\text{s} \mathbf{s} +\mathbf{b}_\text{s} ,
\end{equation}
where $\mathbf{W}_\text{s} \in \mathbb{R}^{|\mathcal{D}|\times u}$ and $\mathbf{b}_\text{s} \in \mathbb{R}^{|\mathcal{D}|}$ are parameters to be learnt. 

The goal of latent space disentanglement is to minimize the overlap of information among latent units in the vector. There are many methods that perform latent space disentanglement using regularization terms to minimize the mutual information between latent units~\cite{chen2018isolating,burgess2018understanding,sanchez2020learning}. For our task, we aim to disentangle the projected text representation such that each unit in the latent vector represents a type of hateful meme targets. Noted that we assume that each meme is likely to concentrate on a certain category of hate (i.e., religion, gender, race, etc.) in most cases. To achieve this objective, we adopt a similar approach to~\cite{ma2020disentangled}, where we maximize the likelihood of the target present in the latent text representation while minimizing for the absent targets. Such a discontinuous $\arg\max$ operation can be fulfilled by applying \textit{Straight-Through Gumbel-Softmax} (STGS) function~\cite{jang2016categorical} over $\mathbf{s}_\text{p}$. Specifically, a continuous vector $\mathbf{z} \in \mathbb{R}^{|\mathcal{D}|}$ is first sampled from the Gumbel-Softmax distribution based on $\mathbf{s}_\text{p}$:


\begin{align}
\label{eq:gumbel}
    u_k &\sim \text{Uniform} (0,1), \\
    g_k &=-\text{log}(-\text{log}(u_k)),\\
    z_k &=\frac{\text{exp}(\text{log}(s_\text{p}^k)+g_k)/\tau}{\sum_{k=1}^{|\mathcal{|D|}|}\text{exp}(\text{log}(s_\text{p})^k+g_k)/\tau},
\end{align}
where $s_\text{p}^k$ is the $k$-th element in $\mathbf{s}_\text{p}$. In the forward pass, the STGS function then transforms the continuous vector sampled from the Gumbel-Softmax distribution into a one-hot vector~\cite{jang2016categorical}:

\begin{equation}
    l_\text{s}^k = \left \{ 
    \begin{array}{l}
    1\quad k=\arg\underset{m}{\min} z_m \\
    0 \quad else
    \end{array}
    \right .
\end{equation}
Finally, the exploitation of STGS to generate the disentangled text representation can be simplified as:
\begin{equation}
    \mathbf{l}_\text{s} =\text{Gumbel-Softmax} (\mathbf{s}_\text{p}),
\end{equation}
where $\mathbf{l}_\text{s}=\{s_\text{p}^k\}_{k=1}^{|\mathcal{D}|}$ as generated by Equation~\ref{eq:gumbel}. For example, in Figure~\ref{fig:arch}, the disentangled text representation would be a one-hot latent vector where 1 is assigned to the latent unit that represents `woman' (i.e., target). The disentangled text representation $\mathbf{l}_\text{s}$ will be used in the self-supervised matching with the visual disentangled latent representation in section~\ref{sec:visdis}. 

Via learning a disentangled text representation in the projected latent space, we update the text representation $\mathbf{s}$ through the back-propagation process, thus fine-tuning $\mathbf{s}$ to be more representative of the target information. We then use the updated $\mathbf{s}$ for hateful meme classification, as discussed in Section~\ref{sec:classification}.

\subsection{Visual Representation Learning}
\label{sec:visual}
After learning the text representation, we focus on learning the disentangled latent representation in the visual modality. The input of this module is the image features pre-processed using Faster R-CNN~\cite{ren2016faster}. Formally, we define the set of image features as $\mathbf{V}=\{\mathbf{v}_i\}_{i=1}^N$, where $\mathbf{v}_i \in \mathbb{R}^d$ is the feature for the $i$-th detected region using \textit{Faster R-CNN}~\cite{ren2016faster} and $N$ is the number of detected region.

\textbf{Attention-Based Image Encoder}
To enable better interaction between visual and textual modality, we adopted the multi-head attention proposed in~\cite{attentionashish} to learn an attended latent visual representation of the meme. Specifically, we leverage the textual representations $\textbf{C}$ generated by BERT text encoder as guidance to attend the feature map $\mathbf{V}$ and generate the attended visual representation using the attention-based image encoder. The attended visual representation is computed as follow:

\begin{align}
    \label{eq:second} \mathbf{F}_t &=\text{softmax} (\frac{\mathbf{W}_{\text{q},t}\mathbf{C}(\mathbf{W}_{\text{k},t}\mathbf{V})^T}{\sqrt{q}})\mathbf{\mathbf{W}_{\text{v},t}V} \\
    \tilde{\mathbf{F}} &= \text{Concate}(\{\mathbf{F}_t\}_{t=1}^{q}) \\
    \label{eq:fourth}\mathbf{G} &= \mathbf{W}_{\text{f},1}(\text{Relu}(\mathbf{W}_{\text{f},2}\tilde{\mathbf{F}})+\mathbf{b}_{\text{f},2})+\mathbf{b}_{\text{f},1}\\
    \label{eq:fifth} \tilde{\mathbf{V}} &= \tilde{\mathbf{F}} + \mathbf{G} \\
    \mathbf{v}_\text{att} &= \sum_{m=1}^{M} \tilde{\mathbf{v}}_m,
\end{align}
where $q$ denotes the number of times the dot-attention is computed in the multi-head attention. Specifically, the $t$-th attended image feature $\mathbf{F}_t \in \mathbb{R}^{\frac{u}{q}\times M}$ is generated as shown in Equation~\ref{eq:second}, where $\mathbf{W}_{\text{c},t} \in \mathbb{R}^{\frac{u}{q}\times u}$, $\mathbf{W}_{\text{k},t} \in \mathbb{R}^{\frac{u}{q}\times d}$ and $\mathbf{W}_{\text{v},t} \in \mathbb{R}^{\frac{u}{q}\times d}$ are parameters involved in the $t-th$ computation. The attended image features in different aspects are concatenated in row and generate $\tilde{\mathbf{F}} \in \mathbb{R}^{u\times M}$. Similar to ~\cite{attentionashish}, a residual connection is applied to the attended image features as illustrated in Equation~\ref{eq:fourth} and~\ref{eq:fifth}. Finally, weighted average pool over the attended image features $\tilde{\mathbf{V}}$ results in the attended latent visual representation $\mathbf{v}_\text{att}$.

\textbf{Visual-Target Disentanglement}
\label{sec:visdis}
Although the attended visual representation is generated with an attention mechanism, there is no explicit guidance or supervision signal that forces the model to focus on the image regions that are more relevant to the contextual information (i.e., target entities of hateful memes). For instance, in Figure~\ref{fig:arch}, the visual representation should ideally focus on the image region with the three women and to be aware that the focused region infers the ``gender'' as the target of the hateful meme. To make the visual representation more ``target-aware'', we design a latent space matching mechanism, which aims to disentangle the target information from the visual representation and constraint the disentangled visual representation to be consistent with the disentangled latent text representation. 


Similar to the text-target disentanglement, we first project the visual representation $\mathbf{v}_\text{att}$ into the latent space with a linear projection layer:
\begin{equation}
    \mathbf{v}_\text{p}=\mathbf{W}_\text{a}\mathbf{v}_\text{att} + \mathbf{b}_\text{a},
\end{equation}
where $\mathbf{W}_\text{a} \in \mathbb{R}^{|\mathcal{D}|\times u}$ and $\mathbf{b}_\text{a} \in \mathbb{R}^{|\mathcal{D}|}$ are parameters to be learnt.

To disentangle the target information in the visual representation, we introduce a matching loss that constraints $\mathbf{v}_\text{p}$ to be similar to the disentangled text representation $\mathbf{l}_\text{s}$:
\begin{equation}
\label{eq:matching}
    \mathcal{L}_\text{Match} = \sum_{k=1}^{|\mathcal{D}|} l_\text{s}^k log(v_\text{p}^k)+(1-l_\text{s}^k )log(1-v_\text{p}^k),
\end{equation}
where $v_\text{p}^k$ is the $k$-th unit in the visual latent representation $\mathbf{v}_\text{p}$. The matching loss serves as a supervision signal to disentangle the target information in the latent visual representation. The underlying intuition is that the disentangled text representation $\mathbf{l}_\text{s}$ is trained to disentangle the target in the textual information, and constraining the disentangled visual representation to be similar to the disentangled text representation would enable the target latent unit to be maximized in disentangled visual representation. For example, in Figure~\ref{fig:arch}, the matching loss will take guidance from the disentangled text representation and enforce the disentangled visual representation to maximize the latent unit representing ``gender''. 

Similarly to the text representation learning, the visual representation $\mathbf{\mathbf{v}_\text{att}}$ would be updated through the back-propagation process, fine-tuning $\mathbf{v}_\text{att}$ to be more representative of the target information. The updated $\mathbf{v}_\text{att}$ would be used for the hateful meme classification discussed in Section~\ref{sec:classification}.


\subsection{Classification}
\label{sec:classification}
To perform hateful meme classification, we leverage a regression layer to generate a \textit{hateful score}. Specifically, if the hateful score is above a threshold, it will be regarded as hateful, otherwise non-hate. By learning the disentangled textual and visual representation and minimizing the matching loss between them in the latent space, the sentence representation $\mathbf{s}$ and the attended image feature $\mathbf{v}_\text{att}$ are both fine-tuned to be more representative of the target information. Finally, we concatenate $\mathbf{s}$ and $\mathbf{v}_\text{att}$, and feed the concatenated representation to a regression layer for the score prediction:
\begin{equation}
    y=\sigma ( \mathbf{w}_\text{r}^T[\mathbf{s},\mathbf{v}_\text{att}] + b_\text{r}),
\end{equation}
where $\mathbf{w}_\text{r} \in \mathbb{R}^{2u}$ and $b_\text{r} \in \mathbb{R}$ are parameters to be learnt. Following~\cite{kiela2020hateful}, we set the threshold $\lambda$ as $0.5$: if the score $y$ is above the threshold, it will be predicted as a hateful meme, otherwise, non-hate. 

\textbf{Loss Function}. We optimize the following loss when training our model using mini-batch gradient descent:
\begin{equation}
    \mathcal{L}_{\bm{\theta}} =  \mathcal{L}_{\bm{\theta},\text{Predict}} + \mu \mathcal{L}_{\bm{\theta},\text{Match}},
\end{equation}
where $\bm{\theta}$ denotes parameters of the model, $\mathcal{L}_{\bm{\theta},\text{Match}}$ is the matching loss in the disentangled latent space, computed from the sum of matching loss over all training samples (see Equation~\ref{eq:matching}); and $\mathcal{L}_{\bm{\theta},\text{Predict}}$ is the loss from prediction and $\mu$ is the hyper-parameter balancing the relative importance of both loss types. 
The prediction loss is defined as:
\begin{equation}
   \mathcal{L}_{\bm{\theta},\text{Predict}} = \sum_{s=1}^{|\mathcal{T}|} \hat{y}_s log(y_s) + (1-\hat{y}_s)log(1-y_s),  
\end{equation}
where $\mathcal{T}$ is the training set and $\hat{y}_s$ is the ground-truth label and $y_s$ is the predicted score of the $s$-th training sample. 

\subsection{Implementation Details}
\label{sec:details}
We set the dimension size to 2048 for image features extracted from Faster R-CNN~\cite{ren2016faster}. All textual information is tokenized using BERT~\cite{devlin2018bert} default tokenizer. The length of the sequence is set to $64$, and shorter sentences are padded while longer sentences are truncated. Weights in the BERT-based model are used to initialize the text encoder. The hyperparameter $\mu$ for balancing prediction loss and matching loss is set to be $0.05$ on the FHM dataset and $0.03$ on the MultiOFF dataset. All implementations are done in Pytorch, and we adopt AdamW~\cite{loshchilov2018fixing} as the optimization algorithm to train our model. The size of the minibatch is set to $64$.  

\section{Experiment}
In this section, we will first describe the settings of experiments conducted to evaluate our \textsf{DisMultiHate} model. Next, we discuss the experiment results and evaluate how \textsf{DisMultiHate} fare against other state-of-the-art baselines. We also conduct case studies to qualitatively analyze \textsf{DisMultiHate}'s ability to identify the targets of hateful memes. Finally, we perform error analysis and discuss the limitations of the \textsf{DisMultiHate} model.

\subsection{Evaluation Setting}
\textbf{Dataset.} We evaluate \textsf{DisMultiHate} and the baseline models on the two publicly available hateful memes classification datasets presented in Section~\ref{sec:dataset}. The train-validation-test split of the dataset are illustrated in Table~\ref{tab:dataset}.

\textbf{Evaluation Metrics.} We adopt the evaluation metrics proposed in the hateful meme dataset papers~\cite{kiela2020hateful,suryawanshi2020multimodal}. Specifically, for the evaluation on FHM dataset~
\cite{kiela2020hateful}, we use Area Under the Receiver Operating Characteristic curve (AUROC) and accuracy score as the evaluation metrics. \citet{suryawanshi2020multimodal} had only reported the F1, precision, and recall on the hateful class when they proposed the MultiOFF dataset~\cite{suryawanshi2020multimodal}. However, we noted that due to class imbalance in hate speech classification, most existing studies \cite{fortuna2018survey,schmidt2017survey} have preferred to use weight metrics to evaluate the classification task. Thus, we adopt weighted F1, weighted precision, and weighted recall as the evaluation metrics for the MultiOFF dataset.

\textbf{Baselines.} We benchmark \textsf{DisMultiHate} against the state-of-the-art multimodal methods that were evaluated on the FHM and MultiOFF datasets. We have also included a unimodal baseline for comparison. Specifically, we applied the pre-trained BERT~\cite{devlin2018bert} to extract text features from the meme's text and feed the extracted text features to a fully connected layer for classification.  

For FHM dataset, we compare with the four best performing multimodal models reported in the original dataset paper~\cite{kiela2020hateful}, namely: \textbf{ViLBERT}~\cite{lu2019vilbert}, \textbf{ViLBERT-CC} (i.e., ViLBERT pre-trained on Conceptual Captions~\cite{sharma2018conceptual}), \textbf{VisualBERT}~\cite{li2019visualbert}, and \textbf{VisualBERT-COCO} (i.e., VisualBERT pre-trained on COCO~\cite{lin2014microsoft}). There are many interesting solutions proposed by the Facebook hateful memes classification challenge participants~\cite{zhu2020enhance,zhou2020multimodal,lippe2020multimodal,zhang2020hateful,muennighoff2020vilio,zhong2020classification}. For our evaluation, we benchmark against the methods explored by the top-performing participant\footnote{https://ai.facebook.com/blog/hateful-memes-challenge-winners/}. Specifically, we benchmark against the methods proposed in Zhu's exploration~\cite{zhu2020enhance}: \textbf{ERNIE-Vil}~\cite{yu2020ernie}, \textbf{UNITER}~\cite{chen2020uniter},  \textbf{VILLNA}~\cite{gan2020large}, and \textbf{VL-BERT}~\cite{su2019vl}. We have reproduced the model using the code\footnote{https://github.com/HimariO/HatefulMemesChallenge} released in \cite{zhu2020enhance}. We also adopt the same data augmentation method proposed in \cite{zhu2020enhance} to enhance the models. Specifically, all the reproduced models are augmented with entity tags retrieved using Google Vision Web Entity Detection, and \textbf{VL-BERT} is also further enhanced with demographic information extracted using FairFace~\cite{karkkainen2019fairface}.

For MultiOFF dataset, we compare with the multimodal methods reported in the dataset paper~\cite{suryawanshi2020multimodal}, namely: \textbf{StackedLSTM+VGG16}, \textbf{BiLSTM+VGG16}, and \textbf{CNNText+VGG16}. For these multimodal baselines, the researchers first extract the image features using VGG16~\cite{simonyan2014very} pre-trained on the ImageNetdataset, and extract text features using various text encoders (e.g., BiLSTM). Subsequently, the extracted image and text features are concatenated before feeding into a classifier for hateful meme classification. As the MultiOFF dataset is relatively new, few studies have benchmarked on this dataset. Therefore, as additional baselines, we reproduced the methods proposed by \citet{zhu2020enhance} on the MultiOFF dataset.

\subsection{Experimental Results}
\begin{table}[t]
  \caption{Experimental results on FHM dataset.}
  \label{tab:fhm-result}
  \begin{tabular}{c|cc}
    \hline
    Model & Acc. & AUROC\\
    \hline
    BERT (unimodal) & 58.3 & 64.7  \\\hline
    ViLBERT & 62.2 & 71.1  \\
    VisualBERT & 62.1 & 70.6 \\
    ViLBERT-CC & 61.4 & 70.1 \\
    VisualBERT-COCO & 65.1 & 74.0\\\hline
    ERNIE-VIL & 69.0 & 78.7 \\
    UNITER & 68.6 & 78.0 \\
    VILLNA & 71.2  & 78.5  \\
    VL-BERT & 71.4 & 78.8  \\\hline
    \textsf{DisMultiHate} (w/o disentangle) & 73.6 & 81.4 \\
    \textsf{DisMultiHate} & \textbf{75.8} & \textbf{82.8} \\\hline
\end{tabular}
\end{table}

\begin{table}[t]
  \caption{Experimental results on MultiOFF dataset.}
  \label{tab:off-result}
  \begin{tabular}{c|ccc}
    \hline
    Model & F1 & Precision & Recall \\
    \hline
    BERT (unimodal) & 56.4 & 56.1 & 57.7  \\\hline
    StackedLSTM+VGG16 & 46.3 & 37.3 & 61.1 \\
    BiLSTM+VGG16 & 48.0 & 48.6 & 58.4 \\
    CNNText+VGG16 & 46.3 & 37.3 & 61.1  \\\hline
    ERNIE-VIL & 53.1 & 54.3 & 63.7  \\
    UNITER & 58.1 & 57.8 & 58.4  \\
    VILLNA & 57.3 & 57.1 & 57.6   \\
    VL-BERT & 58.9 & 59.5 & 58.5   \\\hline
    \textsf{DisMultiHate} (w/o disentangle) &  60.8 & 61.4 & 62.7 \\
    \textsf{DisMultiHate}  & \textbf{64.6} & \textbf{64.5} & \textbf{65.1}   \\\hline
\end{tabular}
\end{table}
We have also included a variant of the \textsf{DisMultiHate}, which performed the hateful meme classification without disentangling the target information. Interestingly, we observe that even without the target disentanglement module, \textsf{DisMultiHate} had outperformed the baselines, demonstrating the strength of our data pre-processing approach on augmenting the model with entity and demographics information. \textsf{DisMultiHate} without target disentanglement has also outperformed the VL-BERT model, which was also augmented with entity and demographics. A possible reason for the performance could be due to \textsf{DisMultiHate}'s ability to learn better textual and visual representations for hateful meme classification. Specifically, the visual representation was attended with the textual information, thereby enhancing the visual features with some form of contextual information. Nevertheless, we noted that the performance of target disentanglement further improves the classification results, suggesting the importance of target information in the hateful meme classification task.

Table~\ref{tab:fhm-result} and \ref{tab:off-result} show the experimental results on the FHM and MultiOFF datasets respectively. In both tables, the highest figures are highlighted in \textbf{bold}. We observed that \textsf{DisMultiHate} outperformed the state-of-the-art baselines in both datasets. \textsf{DisMultiHate} has significantly outperformed the baselines proposed in the original dataset papers. For instance, \textsf{DisMultiHate} has outperformed VisualBERT-COCO by more than 10\% (Acc) on the FHM dataset and outperformed BiLSTM+VGG16 by more than 16\% (F1) on the MultiOFF dataset. \textsf{DisMultiHate} has also outperformed the best baseline by 4\% (Acc) and 5\% (F1) on the FHM and MultiOFF, respectively. We noted that the multimodal methods had outperformed the BERT unimodal baselines in the FHM dataset. However, for the experiment on MultiOFF dataset, we observe that the BERT unimodal baseline is able to achieve competitive performance and outperformed the multimodal baselines proposed in the dataset paper~\cite{suryawanshi2020multimodal}. A possible explanation could be the caption and text information in the MultiOFF memes already contain hateful content. Thus, the unimodal baseline using textual features is able to achieve good performance.

\textbf{Ablation Study}. We also conduct an ablation study to examine the usefulness of entity and demographic information augmented in our \textsf{DisMultiHate} method. Table~\ref{tab:fhm-ablation} and \ref{tab:off-ablation} show the results of the ablation study on FHM and MultiOFF datasets, respectively. We observed that \textsf{DisMultiHate} model augmented with both entity and demographic information had yielded the best performance. 

More interestingly, for the FHM dataset, we observed that without augmenting demographic information yields better performance than without augmenting entity information. However, a different observation is made for \textsf{DisMultiHate} performance on the MultiOFF dataset. Specifically, \textsf{DisMultiHate} not augmented with entity information yields better performance than without augmenting demographic information. The ablation study highlights that the model is highly amenable and adapts to different datasets with varying characteristics.

\begin{table}[t]
  \caption{Ablation study on FHM dataset.}
  \label{tab:fhm-ablation}
  \begin{tabular}{c|cc}
    \hline
    Model & Acc. & AUROC\\\hline
    DisMultiHate (w/o Entity) & 60.6 &68.2  \\
    DisMultiHate (w/o Demo) & 72.8 & 80.8  \\
    DisMultiHate (w/o Entity+Demo) & 62.0  & 70.3 \\\hline
    DisMultiHate  & 75.8 & 82.8 \\\hline
\end{tabular}
\end{table}

\begin{table}[t]
  \caption{Ablation study on MultiOFF dataset.}
  \label{tab:off-ablation}
  \begin{tabular}{c|ccc}
    \hline
    Model & F1 & Precision & Recall \\\hline
    DisMultiHate (w/o Entity) & 62.0 & 64.0 & 63.8\\
    DisMultiHate (w/o Demo) & 60.5& 61.0&61.1 \\
    DisMultiHate (w/o Entity+Demo) & 62.0&62.4 &63.1 \\\hline
    DisMultiHate & 64.6 & 64.5 & 65.1   \\\hline
\end{tabular}
\end{table}



\subsection{Case Study}
The ability to disentangle target information in memes is a core contribution in our \textsf{DisMultiHate} model, and we aim to evaluate this aspect of the model qualitatively. Working towards this evaluation goal, we design an experiment to retrieve relevant memes for a given target query. Specifically, for a given text query (e.g., ``woman''), we first generate its disentangled latent text representation, $\mathbf{l}_\text{q}$, using the process described in Section~\ref{sec:text}. Next, we compute the cosine similarity between $\mathbf{l}_\text{q}$ and the disentangled latent visual representation $\mathbf{v}_p$ of all memes in the FHM dataset. Finally, we retrieved the top $k$ memes with the highest similarity scores with the given target query. Intuitively, if \textsf{DisMultiHate} model is able to disentangle the target information in memes, we should be able to infer the query target from the retrieved memes qualitatively. For example, given the text query ``woman'', we should expect the top $k$ retrieved memes to include woman-related memes.

\begin{table*}[t!]
  \centering
  \caption{Retrieved memes from FHM dataset for a given target. The headers indicate the correctly predicted labels of the retrieved memes.}
  \label{tab:case}
  \begin{tabular}{|c|c|c|c|c| }
    \hline
    \textbf{Target} & \multicolumn{2}{c|}{\textbf{Hateful}} & \multicolumn{2}{c|}{\textbf{Non-Hateful}} \\\hline
    Woman & \begin{minipage}[b]{0.3\columnwidth}
		\centering
		\raisebox{-.5\height}{\includegraphics[width=\linewidth]{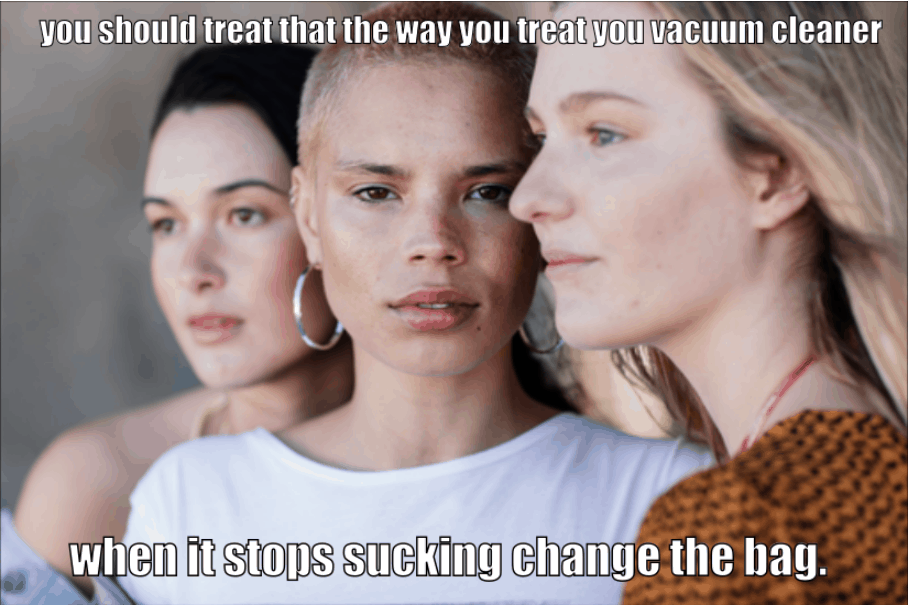}}
	\end{minipage}  &
    \begin{minipage}[b]{0.35\columnwidth}
		\centering
		\raisebox{-.5\height}{\includegraphics[width=\linewidth]{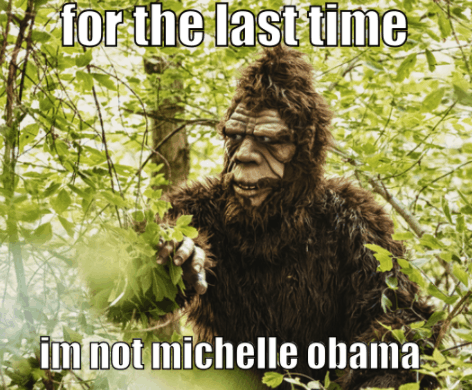}}
	\end{minipage} &
    \begin{minipage}[b]{0.35\columnwidth}
		\centering
		\raisebox{-.5\height}{\includegraphics[width=\linewidth]{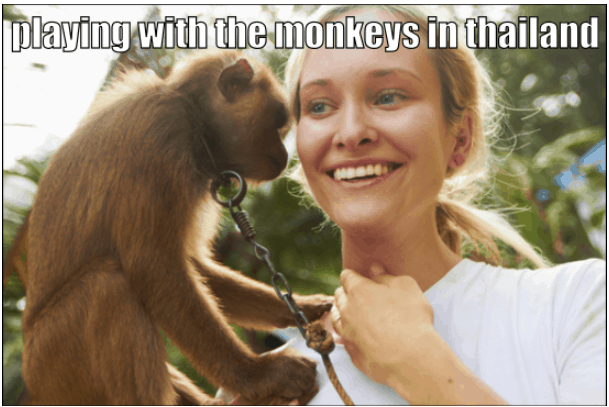}}
	\end{minipage} & \begin{minipage}[b]{0.35\columnwidth}
		\centering
		\raisebox{-.5\height}{\includegraphics[width=\linewidth]{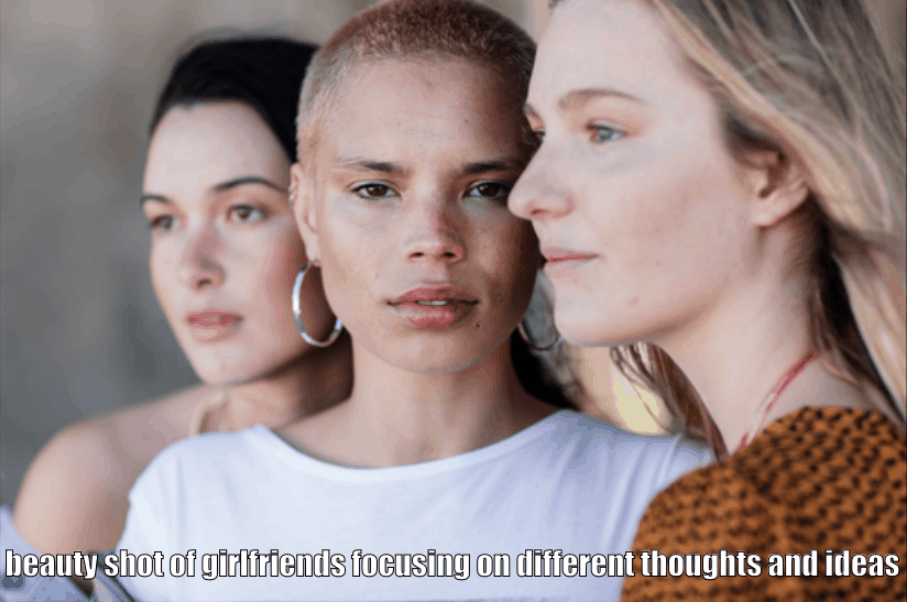}}
	\end{minipage}\\\hline
	Black Man & \begin{minipage}[b]{0.35\columnwidth}
		\centering
		\raisebox{-.5\height}{\includegraphics[width=\linewidth]{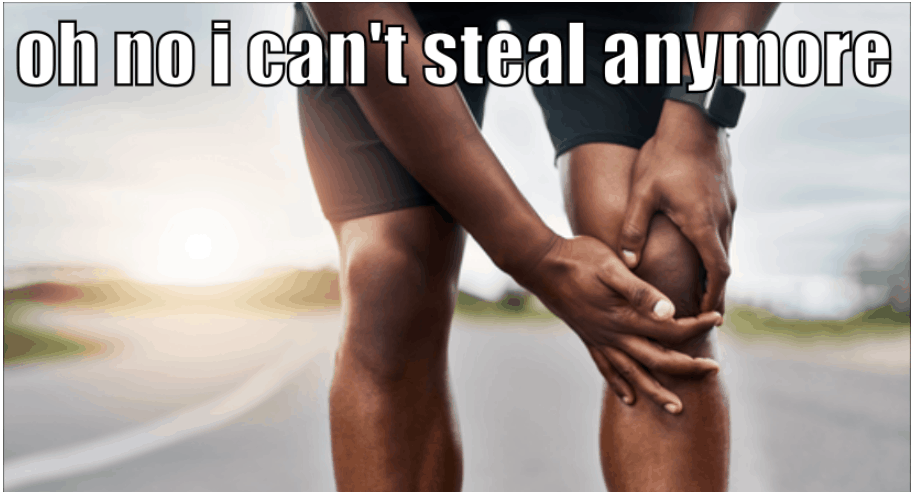}}
	\end{minipage}  &
    \begin{minipage}[b]{0.35\columnwidth}
		\centering
		\raisebox{-.5\height}{\includegraphics[width=\linewidth]{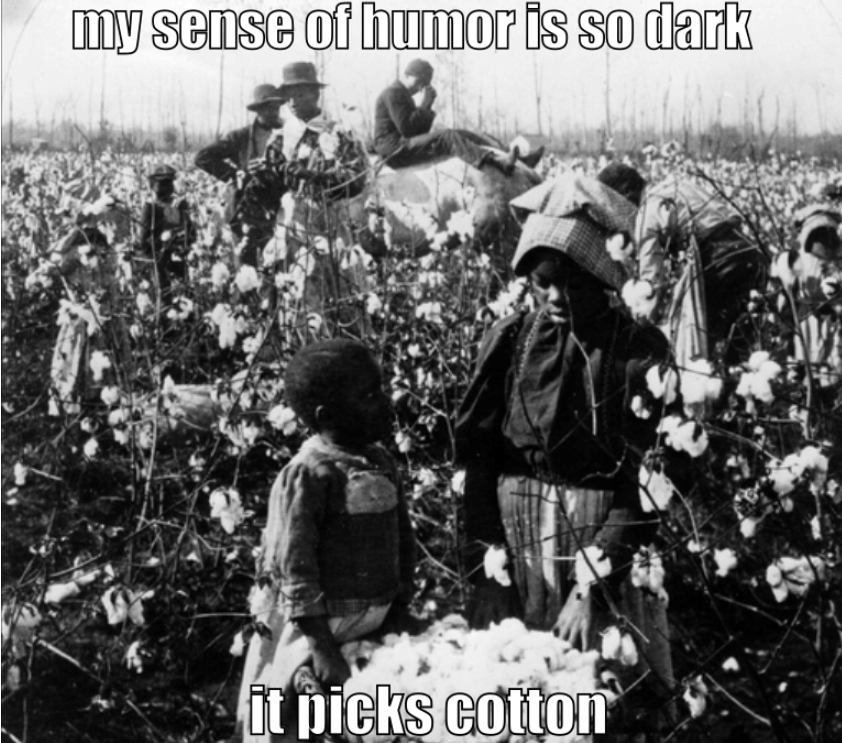}}
	\end{minipage} &
    \begin{minipage}[b]{0.35\columnwidth}
		\centering
		\raisebox{-.5\height}{\includegraphics[width=\linewidth]{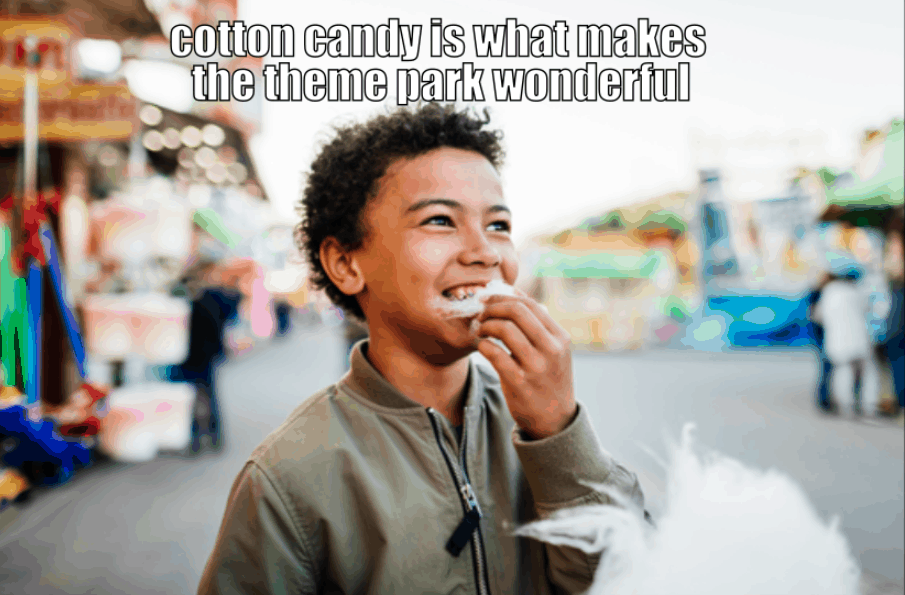}}
	\end{minipage} & \begin{minipage}[b]{0.35\columnwidth}
		\centering
		\raisebox{-.5\height}{\includegraphics[width=\linewidth]{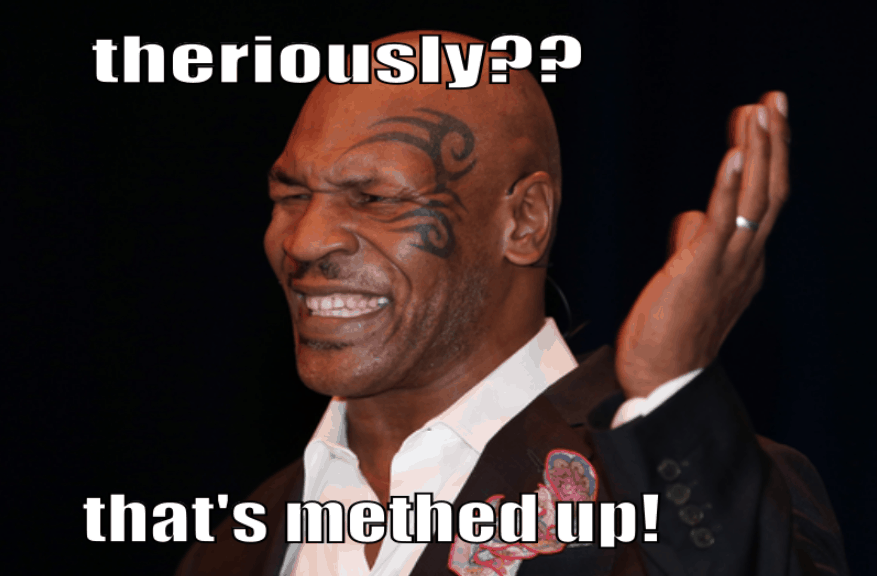}}
	\end{minipage}\\\hline
    \end{tabular}
\end{table*}

\begin{table*}[h]
  \centering
  \caption{Error analysis of wrongly classified memes from FHM dataset.}
  \label{tab:error}
  \begin{tabular}{|c|c|c|c|c| }
    \hline
    \textbf{Meme} & \begin{minipage}[b]{0.35\columnwidth}
		\centering
		\raisebox{-.5\height}{\includegraphics[width=\linewidth]{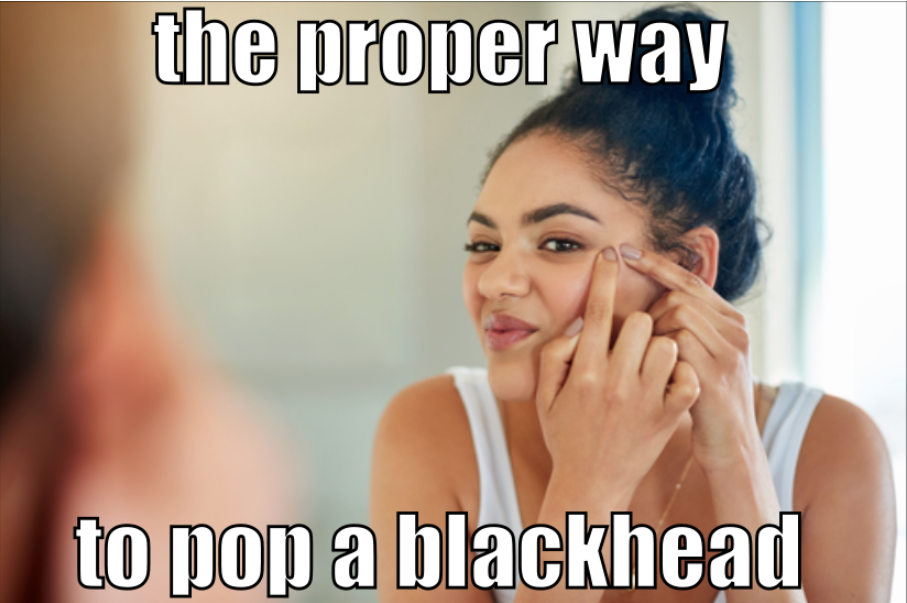}}
	\end{minipage} &
    \begin{minipage}[b]{0.35\columnwidth}
		\centering
		\raisebox{-.5\height}{\includegraphics[width=\linewidth]{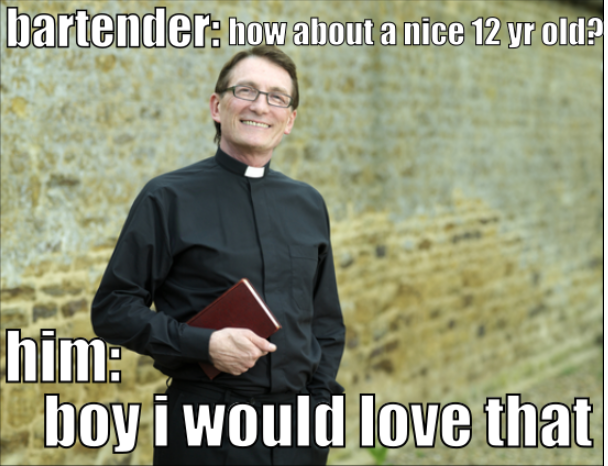}}
	\end{minipage} &
    \begin{minipage}[b]{0.35\columnwidth}
		\centering
		\raisebox{-.5\height}{\includegraphics[width=\linewidth]{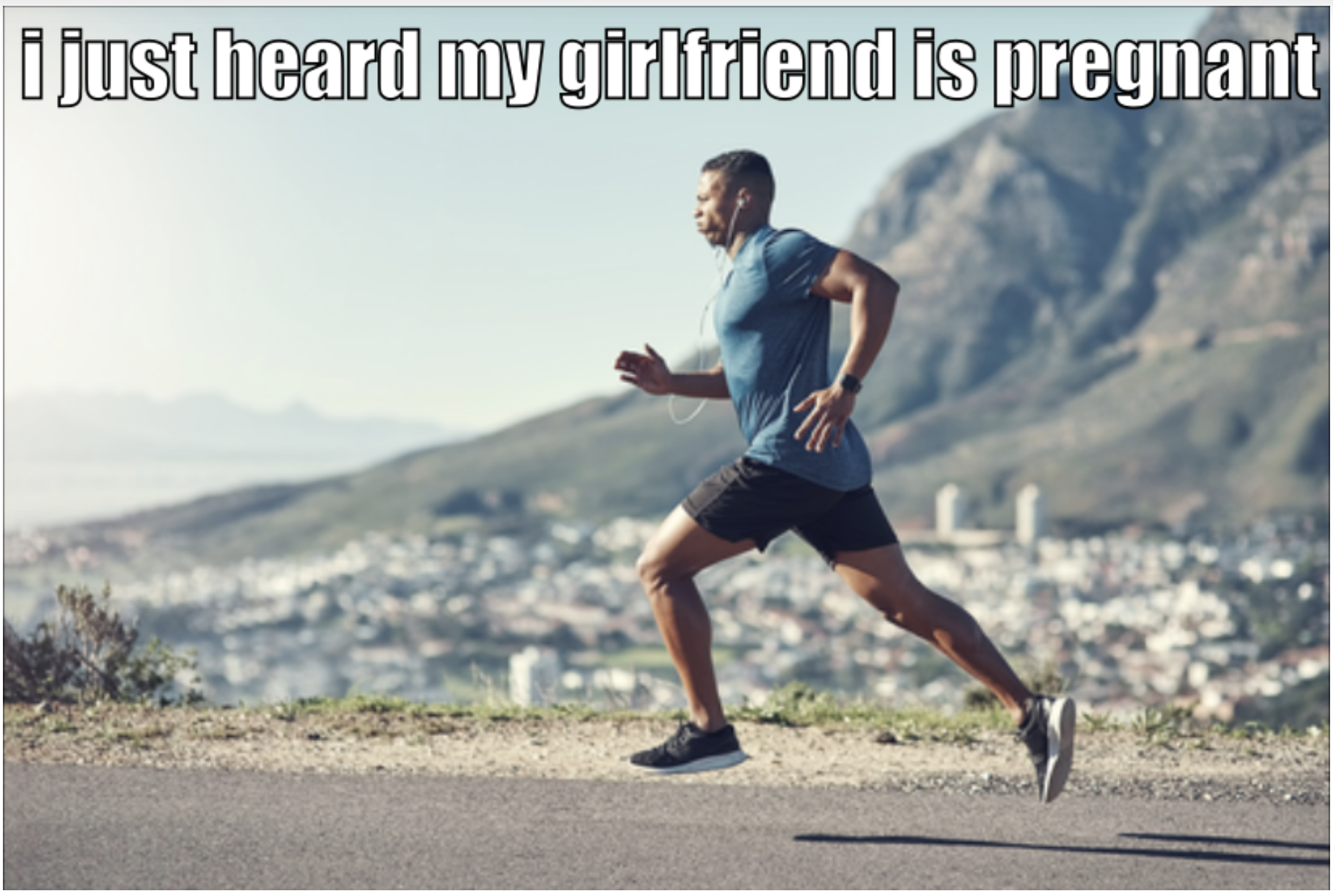}}
	\end{minipage} &
    \begin{minipage}[b]{0.35\columnwidth}
		\centering
		\raisebox{-.5\height}{\includegraphics[width=\linewidth]{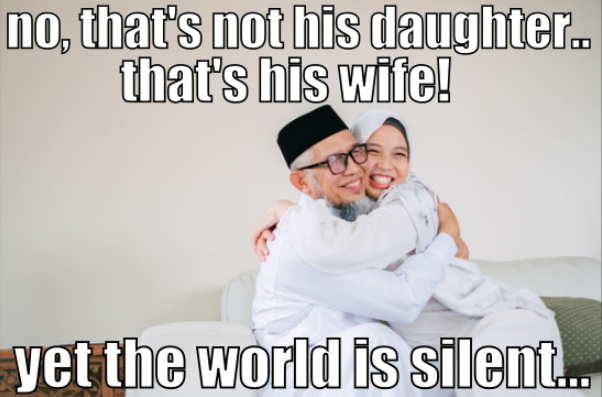}}
	\end{minipage}\\\hline
    \textbf{Actual Label} & Non-Hateful & Non-Hateful & Hateful & Hateful\\\hline
    \textbf{Predicted Label} & \color{red} Hateful & \color{red} Hateful & \color{red} Non-Hateful & \color{red} Non-Hateful\\\hline
    \textbf{Disentangled Target} & Woman & Catholics & Black Man & Muslim Woman\\\hline
    \end{tabular}
\end{table*}

Table~\ref{tab:case} shows the retrieved memes for a given target. Specifically, we retrieve the two most relevant hateful and non-hateful memes for the given target query. We can intuitively infer that the retrieved memes are relevant to the given query. For example, the retrieved memes for the query ``Muslim Woman'' are observed to contain Muslim women in hajib. Interestingly, for the query ``Black Man'', we observe that the second meme is retrieved even though the image is in black and white, and it is not obvious that there are African Americans in the image. However, \textsf{DisMultiHate} is still able to disentangle the ``Black Man'' target in the meme by using relevant textual information such as ``\textit{dark}'' and ``\textit{pick cotton}'' to infer contextual information on the slavery of African American. A similar observation is observed for the ``Woman'' target query, where the second meme does not contain any image of a woman but an ape. However, the second meme is also relevant to the target query as \textsf{DisMultiHate} disentangle the ``Woman'' target in the meme by using relevant textual information such as ``\textit{Michelle Obama}'' to infer contextual information on insulting the individual's physical appearance(i.e., a woman) with a picture of an ape. In summary, the case studies presented in Table~\ref{tab:case} has demonstrated \textsf{DisMultiHate}'s ability to disentangle the target in memes using a combination of textual and visual information captured in the memes. Similar observations were also made for other potential hate speech target queries (e.g., Hispanic, Asian, transgender, etc.).

\subsection{Error Analysis}
Besides analyzing \textsf{DisMultiHate} quantitatively performance over the state-of-the-art baselines, we are also interested in examining the classification errors of \textsf{DisMultiHate}. Table~\ref{tab:error} illustrates four selected examples of \textsf{DisMultiHate}'s wrongly classified memes. For example, \textsf{DisMultiHate} has classified the first meme to be hateful when the actual label of the meme is non-hateful. A possible reason for this error could be the mention of the keyword ``black'' and the disentangled target being ``woman'', which misled the model to make a wrong prediction. 

Our error analysis also reveals some issues with the FHM dataset. For instance, the second meme is annotated as non-hateful in the dataset. However, upon closer examination of the meme, we could infer some form of discrimination towards the Catholics and Priest, and our \textsf{DisMultiHate} has predicted the meme to be hateful. Another issue of the FHM dataset is the potential noise in the dataset. For example, \textsf{DisMultiHate} has wrongly classified the meme as non-hateful when the content is obviously communicating otherwise. We have checked the FHM dataset and found similar memes (i.e., a meme with a running black man) annotated as non-hateful.

\textsf{DisMultiHate} has also wrongly predicted the last meme to be non-hateful as none of the textual keyword, or image features provided the context information that it is hateful. Some form of advance reasoning would be required to understand the hateful context presented in this meme. We could explore adding advanced reasoning modules to classify such memes that require deeper reasoning for future work.

\section{Conclusion}
In this paper, we proposed a novel framework, \textsf{DisMultiHate}, which learns and disentangles the representations of hate speech-related target entities, such as race and gender, in memes to improve the hateful content classification. We evaluated \textsf{DisMultiHate} on two publicly available datasets, and our extensive experiments have shown that \textsf{DisMultiHate} outperformed the state-of-the-art baselines. We have conducted case studies to empirically demonstrated \textsf{DisMultiHate}'s ability to disentangle target information in the memes. We have also performed error analysis and discussed some of the limitations of the \textsf{DisMultiHate} model. We will incorporate a more advanced reasoning module in the model for future works and test the model on more hateful meme datasets. Through applying \textsf{DisMultiHate} to disentangle the target in hateful memes, we also hope to raise awareness of the vulnerable groups targeted in hate speeches in real-world datasets.  


\section*{Acknowledgement}
This research is supported by the National Research Foundation, Singapore under its Strategic Capabilities Research Centres Funding Initiative. Any opinions, findings and conclusions or recommendations expressed in this material are those of the author(s) and do not reflect the views of National Research Foundation, Singapore.

\bibliographystyle{ACM-Reference-Format}
\balance
\bibliography{Main}

\end{document}